\documentclass[12pt]{article}
\usepackage{euscript,amsmath,amssymb,epsfig,latexsym}
\usepackage{amssymb,amsmath}
\usepackage{hyperref}
\usepackage{multicol}
\usepackage{amsfonts}
\usepackage{longtable}
\usepackage[T2A]{fontenc}
\usepackage[cp1251]{inputenc}

 \newcommand{\be}{\begin{equation}}
\newcommand{\ee}{\end{equation}}
\newcommand{\bea}{\begin{eqnarray}}
\newcommand{\eea}{\end{eqnarray}}

\begin{document}

\title{Remarks on the complete integrability of quantum and classical
	dynamical  systems}
\author{Igor V. Volovich\\
\\
{\it Steklov Mathematical Institute of Russian Academy of Sciences,}\\ {\it Moscow, Russia}\\
{\it volovich@mi-ras.ru}}
\date {~}
\maketitle

\begin{abstract}

	It is noted that the Schr\"{o}dinger equation with any self-adjoint  Hamiltonian  is unitary equivalent to a set of non-interacting classical harmonic oscillators and in this sense  any quantum dynamics is completely integrable. Higher order integrals of motion are presented. That does not mean that we can explicitly compute the time dependence for expectation value of any quantum observable. A similar result is indicated for classical dynamical systems in terms of Koopman's approach. Explicit  transformations of quantum and classical dynamics to the free evolution by using direct methods of scattering theory and wave operators are considered. Examples from classical and quantum mechanics, and also from nonlinear partial differential equations and quantum field theory are discussed. Higher order integrals of motion for the multi-dimensional nonlinear Klein-Gordon and Schr\"{o}dinger equations  are mentioned.

\end{abstract}

\newpage

\section {Introduction}

The problem of integration of Hamiltonian systems and  complete integrability has already been discussed in works of Euler, Bernoulli, Lagrange, and Kovalevskaya on the motion of a rigid body in mechanics. Liouville's theorem states that if a Hamiltonian system with $ n $ degrees of freedom has $ n $ independent integrals in involution, then it can be integrated in quadratures, see \cite {AKN}. Such a system is called completely Liouville integrable; for it there is a canonical change of variables that reduces the Hamilton equations to the action-angle form.

An important modern development in studying of integrable systems started with the celebrated work of Gardner, Green, Kruskal and Miura   \cite{GGKM} where the inverse scattering method for solving nonlinear Korteweg-de Vries equations was developed. A complete integrability of the corresponding Hamiltonian system was shown in \cite{FZ}. A generalization to the relativistic sin-Gordon equation was obtained in \cite{ZTF}.

A study of similar quantum systems showed that the remarkable properties of the classical scattering in such models - the absence of multiparticles production and conservation of sets of momenta of asymptotic states, which determined the scattering operator (S-matrix) - are preserved in the relativistic sin-Gordon quantum field theory \cite{AK}. Earlier the complete integrability of some  non-relativistic quantum systems was studied by Bethe, Yang, Baxter and others. Completely integrable classical and quantum systems are considered in numerous  works, in particular in \cite {Koz83, NZ,TF,VV,KT,MPW,Koz18,TS, Koz}. Various notions of integrability have been used. In particular, the Lax representation \cite{NZ}, the Yang-Baxter equation \cite{TF}, an explicit computation of the $S$-matrix \cite{AK}, local and non-local conservation laws \cite{VV, KT, MPW, Koz18}.  The complete integrability of the Schr\"{o}dinger equation in $ \mathbb {C} ^ n $ with a non-degenerate spectrum  is proved in \cite{Koz}.

In this paper, we note that the Schr\"{o}dinger equation $i\dot{\psi}=H\psi$ for any quantum system with an arbitrary self-adjoint Hamiltonian $H$ in a separable Hilbert space ${\cal H}$ is unitary equivalent to a set of classical non-interacting one-dimensional harmonic oscillators and, in this sense, is completely integrable. According to the spectral theorem there exists a unitary map $W:{\cal H}\to L^2(X,\mu),$ where $X$ is a measurable space with a measure $\mu$ such that the Schr\"{o}dinger equation is unitary equivalent to a simpler equation $i\dot{\varphi}_x=\omega_x\varphi_x$ on the space $L^2(X,\mu)$,  $\omega$ being  a function $\omega:X\to \mathbb{R}$ and $x\in X$. The last equation, after decomplexification, describes a family of classical non-interacting harmonic oscillators. By using this isomorphism we construct higher order integrals of motion for the original Schr\"{o}dinger equation. Of course, the task of how to construct explicitly the Hilbert space $L^2(X,\mu)$ and the unitary transformation $W$ is a difficult separate problem. Also, that does not mean that we can explicitly compute the time dependence for expectation value of any quantum observable $A$ in $L^2(X,\mu)$,\,$(A\varphi (t),\varphi(t))$.

  Then we discuss  applications  of scattering theory (not the inverse scattering  transform method but rather the direct method of the wave operators) to study the integrability of various classical and quantum dynamical systems and the construction of integrals of motion. Let $H$ and $H_0$ be self-adjoint operators on Hilbert space and exist the limits $\lim_{t\to\pm\infty}e^{itH}e^{-itH_0}=\Omega_\pm$ which are called the wave operators. One considers only absolute continuous parts of the Hamiltonians. The wave operators $\Omega_\pm $ obey the intertwining property $H\Omega_\pm=\Omega_\pm H_0$. If  $H_0$ is a simple "free" operator then the existence of the intertwining relation can be used to claim the integrability of the system described by the more complicated Hamiltonian $H$. Even if it is so, still there is a problem of how to compute explicitly the wave operators or the scattering operator ($S$-matrix) $S=\Omega^*_+\Omega_-$ which is a task of primary interest.

Examples from classical and quantum mechanics, and also from non-linear partial differential equations and quantum field theory are discussed. Higher order integrals of motion for the multi-dimensional nonlinear Klein-Gordon and Schr\"{o}dinger equations  are proposed.

Note that there is also another direction in studying properties of quantum systems when one takes the large time limit together with small coupling or small density limit,  see \cite {ALV,APV, FSP}.

In the next section  the finite-dimensional case is considered.  In Sect. 3 the complete integrability of arbitrary quantum dynamics, i.e. the Schr\"{o}dinger equation with an arbitrary self-adjoint Hamiltonian is proved. It follows from the spectral theorem. Integrals of motion are found. Also the complete integrability of classical dynamical systems in the Koopman formalism is mentioned (Remark 4). Applications  of direct methods of scattering theory and wave operators for investigation of integrability of classical and quantum mechanics are presented in Sect. 4 and Sect. 5, respectively. Wave operators and integrability for nonlinear partial differential equations such as nonlinear Klein-Gordon and Schr\"{o}dinger equations are discussed in Sect.6. Finally, in Sect.7 we make some remarks about quantum field theory.

 Note that scattering theory
on p-adic field is considered in \cite{Bur}, that can be used in p-adic mathematical physics \cite{VVZ, 30}.

This preprint is an extended version
 of the paper published in \cite{Vol}.
.
\section {Finite-dimensional Case}

We first consider the case of a finite-dimensional Hilbert space $ {\cal H} = \mathbb {C} ^ n $.

{\bf Theorem 1.} The Schr\"{o}dinger equation $ i \dot {\psi} = H \psi $, where $ H $ is a Hermitian operator in $ \mathbb {C} ^ n, \, \, \psi = \psi (t ) \in \mathbb {C} ^ n $, regarded as a classical Hamiltonian system, with a symplectic structure obtained by decomplexification the Hilbert space,
has $ n $ independent integrals in involution.

{\it Proof.} Let $ \omega_j, \, j = 1, ..., n $ be the eigenvalues of the operator $H$.
By diagonalizing the matrix $ H $ using a unitary transformation, we find that the original Schr\"{o}dinger equation is unitary equivalent to the system of equations $ i \dot {\varphi} _j = \omega_j \varphi_j, $ where $ \varphi_j = \varphi_j (t) \in \mathbb {C}, \, \, j = 1, ..., n. $.

Passing to the real and imaginary parts of the function $ \varphi_j (t) = (q_j (t) + ip_j (t))
/ \sqrt {2} $, this system of equations is rewritten in the form of Hamiltonian equations of the family of classical harmonic oscillators
$$
\dot {q} _j = \omega_j p_j, \, \, \dot {p} _j = - \omega_j q_j.
$$
with the Hamiltonian
$$
H_ {osc} = \sum_ {j = 1} ^ n \frac {1} {2} \omega_j (p_j ^ 2 + q_j ^ 2)=(\psi, H\psi).
$$

Define
$$
I_j (\varphi)= | \varphi_j | ^ 2 = \frac {1} {2} (p_ {j} ^ 2 + q_ {j} ^ 2)
, \, \, j = 1, ..., n. $$
The functions $ I_j $ are integrals of motion, independent and in involution,\
$ \{I_j, I_m \} = 0, \, \, j, m = 1, ..., n. $. The corresponding level manifold has the form of an $ n $-dimensional torus. Note that the Hamiltonian is a linear combination of the integrals of motion:
$$
H_ {osc} = \sum_ {j = 1} ^ n \omega_j I_j.
$$
The theorem is proved.

{\bf Remark 1.} We give another, equivalent way of proving the theorem. Let $ e_1, ..., e_n $ be an orthonormal basis of the eigenvectors of the Hermitian operator $ H $ in $ \mathbb {C} ^ n $, i.e. $ He_j = \omega_je_j. $ Then, writing the function $ \psi = \psi (t) $ in the form $ \psi = \sum_j \psi_je_j $, where $ \psi_j = \psi_j (t) \in \mathbb {C} $, we get that the Schr\"{o}dinger equation $ i \dot {\psi} = H \psi $ takes the form $ i \dot {\psi} _j = \omega_j \psi_j. $ Passing to the real and imaginary parts of the function $ \psi_j $ as done above, $ \psi_j = (\xi_j + i \eta_j) / \sqrt {2} $ we get a set of harmonic oscillators with the Hamiltonian $H_{osc}=\sum \omega_j (\eta_j^2 +\xi_j^2)/2$. Note that $H_{osc}$ is equal to the mean value of the quantum Hamiltonian, $H_{osc}=(\psi, H\psi).$ The integrals of motion are defined as follows: $ J_k (\psi) = | \psi_k | ^ 2 = (\eta_k ^ 2 + \xi_k ^ 2) / 2, \, k = 1, ..., n. $ We also add that the solution of the Schr\"{o}dinger equation has the well known form
$$
\psi (t) = \sum_ {j = 1} ^ n \psi_j (0) e ^ {- i \omega_j t} e_j.
$$

\section {Complete integrability of the Schr\"{o}dinger equation}
Let us prove the complete integrability of the Schr\"{o}dinger equation with an arbitrary Hamiltonian.
The Schr\"{o}dinger equation
is completely integrable in the sense that its solutions are unitary equivalent to the complexified solutions of the Hamilton equations for a family of classical harmonic non-interacting oscillators. The Schr\"{o}dinger equation after the unitary transformation is rewritten in the form of equations for the family of classical harmonic oscillators.
This is a consequence of the spectral theorem.

{\bf Theorem 2.} Let $ {\cal H} $ be a separable Hilbert space and $ H $ a self-adjoint operator with a dense domain $ D (H) \subset {\cal H} $.
Then the Schr\"{o}dinger equation
$$
i \frac {\partial } {\partial t}\,\psi (t) = H \psi (t), \, \, \, \psi (t) \in D (H), \, \, t \in \mathbb {R}
$$
completely integrable in the sense that this equation is unitary equivalent to the
complexified system of  equations for a family of classical non-interacting harmonic oscillators. There exists a set of non-trivial integrals of motion for the arbitrary Schr\"{o}dinger equation.

{\it Proof.}
The solution of the
Cauchy problem for the Schr\"{o}dinger equation
$$
i \frac {\partial } {\partial t}\,\psi (t) = H \psi (t), \, \, \, \psi (0) = \psi_0 \in D (H)
$$
by the Stone theorem has the form $ \psi (t) = U_t \psi_0 $, where $ U_t = e ^ {- itH} $ is the group of unitary operators,
$ t \in \mathbb {R}. $ Then, by the spectral theorem \cite{DS, RS}, there exists a measurable space
$ (X, \Sigma) $ with $\sigma$-finite measure $ \mu $, and a measurable finite a.e. function $ \omega: X \to \mathbb {R} $ such that
there is a unitary transformation $ W: {\cal H} \to L ^ 2 (X, \mu) $ such that $ U_t = W ^ * U_t^{(0)} W $ where $U_t^{(0)}=e ^ {- itM_ \omega}$, where $ M_ \omega $ is the operator of multiplication by the function $ \omega $.
On the corresponding domain one has
$ H = W ^ * M_ \omega W $.

The initial Schr\"{o}dinger equation in $ {\cal H} $ goes over under the unitary transformation of $ W $ into the Schr\"{o}dinger equation in $ L ^ 2 (X, \mu) $ of the form
$$
i \frac {\partial } {\partial t} \,\varphi_x (t)= \omega_x \varphi_x (t), \, \, x \in X
$$

Passing to the real and imaginary parts of the function $$ \varphi_x (t) = \frac {1} {\sqrt {2}} (q_x (t) + ip_x (t)), $$ this Schrodinger equation is rewritten in the form of equations of the family of classical harmonic oscillators
$$
\dot {q} _x = \omega_x p_x, \, \, \dot {p} _x = - \omega_x q_x, \, \, x \in X.
$$

These equations can be obtained from the Hamiltonian \footnote{I am grateful to B.~O.~Volkov for discussion. For considerations of infinite-dimensional Hamiltonian systems see \cite{ CM, KS, Kuk} and refs therein.}
$$
H_{osc}=\int_xH_xd\mu,\,\,\,H_x=\frac{1}{2}\omega_x(p_x^2+q_x^2),
$$
by using the relation
$$
\delta H_{osc}=\int_X\delta H_x d\mu=\int_X(\dot{q}_x\delta p_x -\dot{p}_x\delta q_x)d\mu.
$$
Note also that one has $H_{osc}=(\omega\varphi,\varphi)_{L^2}=(H\psi),\psi)_{{\cal H}}$.

{\it Integrals of motion.}

{\bf Lemma 1.} Let $I:L^2(X)\to \mathbb {R}$ be an integral of motion for  the dynamics $U_t^{(0)}, \, I(U_t^{(0)}\varphi)=I(\varphi),\,\varphi\in L^2(X)$.
Then $J:{\cal H}\to \mathbb {R},$ where $J(\psi)=I(W\psi),\psi\in{\cal H},$
is the integral of motion for dynamics $U_t,\,J(U_t\psi)=J(\psi)$.

Proof of Lemma 1 goes as follows:
$$
J(U_t\psi)=I(WU_t\psi)=I(WU_tW^* W\psi)=I(U_t^{(0)} W\psi)=I(W\psi)=J(\psi).
$$
Now let us consider the following integrals of motion for  the dynamics $U_t^{(0)}$:

$$
I_\gamma=I_\gamma (\varphi) = \int_X \gamma_x | \varphi_x | ^ 2 d\mu =\int_X \gamma_x \frac {1} {2} (p_x ^ 2 + q_x ^ 2)d\mu,
$$
for any $\gamma\in L^\infty (X)$.
Then, by Lemma 1, one gets an integral of motion $J_\gamma$ for dynamics $U_t$.
The theorem is proved.

{\bf Remark 2.} Consider an example when $X=\mathbb{R}$ and  the Hilbert space is $ L ^ 2 (\mathbb {R}, d \mu) $, where $ \mu $ is a
Borel measure on the line. Any measure $ \mu $ on $ \mathbb {R} $
admits a unique expansion in the sum of three measures $ \mu = \mu_ {pp} + \mu_ {ac} + \mu_s $, where $ \mu_ {pp} $ is purely pointwise, $ \mu_ {ac} $ is absolutely continuous with respect to the Lebesgue measure and $ \mu_s $ is continuous and singular with respect to Lebesgue measure, and we have respectively
$$
L ^ 2 (\mathbb {R}, \, d \mu) = L ^ 2 (\mathbb {R}, \, d \mu_ {pp}) \oplus L ^ 2 (\mathbb {R}, \, d \mu_ {ac}) \oplus L ^ 2 (\mathbb {R}, \, d \mu_s).
$$
A purely point measure has the form $ \mu_ {pp} = \sum c_j \delta_ {s_j} $, where $ c_j> 0 $, the sum over $ j $ contains a finite or infinite number of terms, $ s_j $ real numbers and $ \delta_ {s_j} $ Dirac delta function. Then we have
$$
\int | \varphi_x | ^ 2 \, d \mu_ {pp} = \sum_ {j} c_j | \varphi_ {s_j} | ^ 2.
$$
The integrals of motion in this case have the form
$$
I_j(\varphi) = | \varphi_ {s_j} | ^ 2 = \frac {1} {2} (p_ {s_j} ^ 2 + q_ {s_j} ^ 2),
$$
and the Hamiltonian of the system of oscillators is
$$
H_ {osc} = \sum_ {j} \frac {\omega_ {s_j}} {2} (p_ {s_j} ^ 2 + q_ {s_j} ^ 2)
$$
with ordinary Poisson brackets, where sequences are considered for which the series converges.

Similarly, an absolutely continuous measure has the form $ d \mu_ {ac} = f \, dx $, where $ f \geq 0 $ is a locally integrable function and $ dx $ is a Lebesgue measure. Then we have
$$
\int | \varphi_x | ^ 2 \, d \mu_ {ac} = \int  | \varphi_x | ^ 2f_x \, dx.
$$
The integrals of motion in this case have the form
$$
I_\gamma =\int\gamma_x | \varphi_ {x} | ^ 2 f_x dx = \int\gamma_x\frac {1} {2} (p_ {x} ^ 2 + q_ {x} ^ 2)f_x dx,
$$
where $\gamma\in L^\infty$.

One would write formally for the density of the integrals
$$
I_x = | \varphi_ {x} | ^ 2 = \frac {1} {2} (p_ {x} ^ 2 + q_ {x} ^ 2), \, \, x \in supp f,
$$
and the Hamiltonian of the system of oscillators is
$$
H_ {osc} = \int \frac {\omega_ {x}} {2} (p_ {x} ^ 2 + q_ {x} ^ 2) \, f_x dx
$$
with ordinary Poisson brackets, where functions are considered for which the integral converges.

{\bf Remark 3.}  It is known that another version of the spectral theorem, using the direct integral of Hilbert spaces, allows us to determine the invariants of the Hamiltonian with respect to unitary transformations. The Hilbert space $ {\cal H} $ is unitary equivalent to the direct integral of the Hilbert spaces $ \{{\cal H} _ {\lambda}, \, \lambda \in \sigma (H) \} $, where $ \sigma (H ) $ spectrum of $ H $, with respect to some measure $ \nu $ on $ \sigma (H) $,
$$
\int _ {\sigma (H)} ^ {\oplus} {\cal H} _ {\lambda} d \nu (\lambda),
$$
the operator $ H $ in $ {\cal H} _ {\lambda} $ corresponds to multiplication by $ \lambda $. The invariants of $ H $ are the spectrum $ \sigma (H) $, the measure $ \nu $ (modulo absolute continuity) and the dimension of the Hilbert spaces $ {\cal H} _ {\lambda} $.

{\bf Remark 4.} The complete integrability of the Liouville equation in Koopman's approach to classical mechanics and in general any dynamical system is proved in a similar way. Let $ (M, \Sigma, \alpha, \tau_t) $ be a dynamical system, where $ (M, \Sigma) $
measurable space with measure $ \alpha $ and $ \tau_t, \, t \in \mathbb {R} $ group of measure-preserving transformations $ M $. Then the Koopman transform defines a group of unitary operators $ U_t $ in $ L ^ 2 (M, \alpha) $,
$$
(U_tf) (m) =
f (\tau_t (m)), \, f \in L ^ 2 (M, \alpha) $$
Repeating the proof of Theorem 2, we see that the group $ U_t $ is unitary equivalent to the family of harmonic oscillators and, in this sense, any dynamical system is completely integrable.

{\bf Remark 5.}
Solution of the Cauchy problem for the Schr\"{o}dinger equation in $ L ^ 2 (X, \mu) $,
$$
i \frac {\partial} {\partial t} \, \varphi_x (t)= \omega_x \varphi_x (t), \, \, \, \varphi_x (0) = \varphi_ {0x}, \, \, x \in X
$$
has the form
$$
\varphi_x (t) = e ^ {- it \omega_x} \varphi_ {0x}, \, \, x \in X.
$$
One can expect that quantum chaos \cite{QC} is related with ergodicity of classical harmonic oscillators for irrational frequencies $\omega_x$.

{\bf Remark 6.} The above interpretation of the spectral theorem as complete integrability in the sense of reducing dynamics to a set of harmonic oscillators
does not mean that such a task is easy to perform for specific systems. It is interesting to compare the effectiveness (or complexity) of complete integrability in the above sense of the spectral theorem with the efficiency/complexity of constructing solutions of Hamiltonian systems that are completely integrable in the sense of Liouville. Apparently, the change of variables in the Liouville theorem, which reduces the initial dynamics to action-angle variables on a level manifold, is an analogue of the unitary transformation in the spectral theorem, which reduces the original dynamics to a set of harmonic oscillators.

\section {Direct methods of scattering theory and \\ integrability of classical dynamical systems}

The inverse scattering problem method, based on the representation of certain nonlinear equations in the Lax form, is used, as is known, to prove the complete integrability of such equations, usually on a straight line or on a plane, see, for example, \cite{NZ}. Here we look at the use of direct scattering theory methods
to prove the complete integrability and the construction of integrals of motion in the theory of classical and quantum dynamical systems. This will provide  a constructive example to
the above general result on the complete integrability of arbitrary quantum and classical systems. Note that a general method was proposed in \cite{VV} for constructing conserved currents for a large class
of (multidimensional) nonlinear equations by using a dual linear equation.

\subsection {Wave operators and integrals of motion}

 Let $ \phi (t) $ and $ \phi_0 (t) $ be a pair of groups of automorphisms of the phase space in the classical case or groups of unitary transformations in  quantum case, $ t \in \mathbb {R} $. Suppose there exist the limits, called wave operators, see \cite{Sim, DG, RS3, Yaf},
 $$
 \lim_ {t \to \pm \infty} \phi (-t) \phi_0 (t) = \Omega _ {\pm}
 $$
 The wave operators have the properties of intertwining operators; on the appropriate domain, we have
 $$
 \phi (t) \Omega _ {\pm} = \Omega _ {\pm} \phi_0 (t)
 $$
 Thus, the interacting dynamics $ \phi (t) $ is reduced to "free"
$ \phi_0 (t) $. The following form of Lemma 1 holds

{\bf Lemma 1a.} Let $ I_0 (z) $ be the integral of motion for the dynamics of $ \phi_0 (t) $, i.e.
$ I_0 (\phi_0 (t) z) = I_0 (z) $, here $ z $ is a phase space point or a Hilbert space vector. Then $ I (z) = I_0 (\Omega _ {\pm} ^ {- 1} z) $ is the integral of motion for the dynamics of $ \phi (t) $.

Indeed, we have
$$
I (\phi (t) z) = I_0 (\Omega _ {\pm} ^ {- 1} \phi (t) z) = I_0 (\Omega _ {\pm} ^ {- 1} \phi (t) \Omega _ {\pm}
\Omega _ {\pm} ^ {- 1} z) = I_0 (\phi_0 (t)
\Omega _ {\pm} ^ {- 1} z) = I_0 (
\Omega _ {\pm} ^ {- 1} z) = I (z).
$$
 
 \subsection {Particle in a potential field}

Let us consider in more detail the motion of a particle in a potential field. Let $ \Gamma = M \times M ^ {'} $ phase (symplectic) space, where $ M = \mathbb {R}^ n $ is the configuration space and $ M ^ {'} = \mathbb {R} ^ n $ is the dual space of momenta. Hamiltonian has the form
$$
H (x, \xi) = \frac {1} {2} \xi ^ 2 + V (x),
$$
where $ \xi \in M^\prime$ and the function $ V: M \to \mathbb {R} $ is bounded and the force $ F (x) = - \bigtriangledown V (x) $ is locally Lipschitz. Then the solution  $ (x (t, y, \eta), \xi (t, y, \eta)) $ of the equations of motion
$$
\dot {x} (t) = \xi (t), \, \, \,
\dot {\xi} (t) = F (x (t))
$$
with initial data
$$
x (0, y, \eta) = y, \, \, \, \xi (0, y, \eta) = \eta, \, \, y \in M, \, \eta \in M^\prime
$$
exists and is unique for all $ t \in \mathbb {R} $. We will denote $ \phi (t) (y, \eta) = (x (t, y, \eta), \xi (t, y, \eta)). $ The free Hamiltonian is $ H_0 (x, \xi) = \xi ^ 2/2, \, $ the solution of the equations of motion has the form $ \phi_0 (t) (y, \eta) = (y + t \eta, \eta). $

Let the following conditions also be satisfied
$$
\lim_ {| x | \to \infty} V (x) = 0, \, \, \int_0 ^ {\infty} \sup_ {| x | \geq r} | F (x) | dr <\infty
$$
Then for any $ (y, \eta) \in M \times M^\prime $ there is a limit \cite{DG}
$$
\lim_ {t \to \infty} \frac {x (t, y,\eta)} {t} = \xi _ + (y, \eta),
$$
moreover, if $ \xi _ + (y, \eta) \neq 0 $, then there is a limit
$$
\lim_ {t \to \infty} \xi (t, y, \eta) = \xi _ + (y, \eta),
$$
 and the inverse image of $ {\cal D} = \xi _ + ^ {- 1} (M '\setminus \{0 \}) $ is the open set of all paths with positive energy unbounded for $ t \to \infty $.

{\bf Proposition 1} (see Theorem 2.6.3 in \cite{DG}). Let the force satisfy the conditions
$$
\int_0 ^ {\infty} \sup_ {| x | \geq r} | \partial_x ^ {\alpha} F (x) | (1 + r ^ 2) ^ {1/2} dr <\infty, \, \,
|\alpha |= 0,1.
$$
Then there exists the limit
$$
\lim_ {t \to \infty} \phi (-t) \phi_0 (t) = \Omega_ +
$$
uniformly on compact sets in $ M \times (M '\setminus \{0 \}) $. The mapping $ \Omega_ +:
M \times (M '\setminus \{0 \}) \to {\cal D} $ is symplectic, continuous and one-to-one. There are relations
$$
H \, \Omega _ + = H_0,\,\,\,
\phi (t) \, \Omega _ + = \Omega _ + \, \phi_0 (t).
$$

{\bf Theorem 3.} The equations of motion of a particle in a potential field satisfying the above  conditions determine an integrable system in the sense that the dynamics of $ \phi (t) $ is symplectically equivalent to the free dynamics of $ \phi_0 (t) $ on the domain $ { \cal D} \, $ in $ 2n $ - dimensional phase space, $
\Omega _ + ^ {- 1} \phi (t) \, \Omega _ + = \, \phi_0 (t).
$ On $ {\cal D} $ there are $ n $ independent integrals of motion in involution.

{\it Proof} follows from   Proposition 1 and Lemma 1a. Let $ I_ {0, j}: {\cal D} \to \mathbb {R}, \, j = 1,2, ..., n $ be the integral of motion for free dynamics $ \phi_0 (t) $, of the form $ I_ {0, j} (y, \eta) = \eta_j$. Then $ I_j (z) = I_ {0, j} (\Omega _ + ^ {- 1} z) $ is the integral of motion for the dynamics of $ \phi (t). $

\section {Wave operators and integrability in \\ quantum mechanics}

Consider the Schr\"{o}dinger equation in $ L ^ 2 (\mathbb {R} ^ n) $ of the form
\be\label{Schrod}
i \frac {\partial \psi} {\partial t} = H \psi,
\ee
where $ H = H_0 + V (x) $, $ H_0=-\bigtriangleup $ - Laplace operator and the potential $ V $ is short-range, i.e. it satisfIes
$$
|V(x)|\leq C(1+|x|)^{-\nu},\,\,\,x\in\mathbb{R}^n,\,\,\nu>1.
$$
Then $ H $ defines a self-adjoint operator and there exist the wave operators $\Omega_\pm$, which are complete and diagonalize $ H,\,\,
H\Omega_\pm = \Omega_\pm H_0 $ \cite{RS3, Yaf}. So, one can construct higher  order integrals of motion according to the scheme described above.

Note that if $\varphi=\varphi (t,x)$ is a smooth solution to the free Schr\"{o}dinger equation $i\dot{\varphi}+\bigtriangleup\varphi=0$ then its derivatives $\partial _x^\alpha\varphi$ also are solutions of this equation.
Hence, one can get higher integrals of  motion just by replacing $\varphi$ by
$\partial _x^\alpha\varphi$ in the known integral  $I_0(\varphi)=\int_{\mathbb{R}^n} |\varphi|^2 dx $.  If $\varphi$ is in the Sobolev space $H^k(\mathbb{R}^n)$ then one gets a set of integrals of motion: $I_0=\int |\varphi|^2 dx ,\,\,I_\alpha=\int |\partial _x^\alpha\varphi|^2 dx,\,|\alpha|\leq k$. Now, according to Lemma 1a, we obtain integrals of motion $J_0, J_\alpha^\pm (\psi)=I_\alpha (\Omega_\pm^{-1}\psi)$ for the original  Schr\"{o}dinger equation (\ref{Schrod}) with the potential $V(x)$.

Remark also that using the Fourier transform $F\varphi = \tilde{\varphi}$
one can write the free dynamics in the form $U_t^{(0)}\tilde{\varphi}(k)=e^{itk^2}\tilde{\varphi}(k)$ with integrals of motion $I_\gamma(\varphi)=\int_{\mathbb{R}^n}\gamma (k) |\tilde{\varphi}(k)|^2dk,$ where $\gamma\in L^\infty (\mathbb{R}^n) $.

\section{Wave operators and integrability for \\  nonlinear partial differential equations}

Let us consider  a nonlinear Klein - Gordon equation of the form
\be\label{NLKG}
\ddot{u}-a^2\bigtriangleup u+m^2 u+f(u)=0, \,\,t\in\mathbb{R},\,\,x\in\mathbb{R}^n,,
\ee
where $a>0, \, m\geq0.$ 

Eq. (\ref{NLKG}) is a Hamiltonian system with the Hamiltonian
\be\label{HNLKG}
H=\int_{\mathbb{R}^n}(\frac{1}{2})p(x)^2 + \frac{1}{2}(\nabla u)^2 +          \frac{1}{2} m^2u^2 +V(u(x))dx =H_0 +V
\ee
where $V^\prime=f$ and the
the Poisson brackets are $\{p(x), u(y)\}=\delta(x-y)$

One can take $f(u)=\lambda |u|^2 u,\,\lambda \geq 0$ and $n=3,$ more general case is considered in numerous papers, see, for instance \cite{GV, Nak, ZJ} and ref's therein. If the initial data $(u(0),\,\dot{u}_t(0))\in H^1(\mathbb{R}^n)\times L^2(\mathbb{R}^n)$ then there exists a global solution $u(t)$ of Eq (\ref{NLKG}) with these initial data. Here $H^1(\mathbb{R}^n)$ is the Sobolev space.  For any solution of (\ref{NLKG})
there exists a unique pair $(v_+,v_{-})$ of solutions for the free Klein-Gordon equation
\begin{equation}\label{KG}
\ddot{v}-a^2\bigtriangleup v+m^2 v=0,
\end{equation}
such that
$$
\lim_{t\to\pm\infty}\|(u(t),\,\dot{u}(t))-(v_{\pm}(t),\dot{v}_{\pm}(t))  \|_{H^1\times L^2}=0.
$$
Moreover, the correspondence $(u(0),\,\dot{u}(0))\mapsto (v_{\pm}(0),\dot{v}_{\pm}(0))$ defines homeomorphisms (wave operators) $\Omega_\pm$ on $H^1\times L^2$.

{\it Integrals of motion.}

If $v=v(t,x)$ is a smooth solution of Eq. (\ref{KG}) then its partial derivatives
$\partial_x^\alpha v$ is also the solution of Eq. (\ref{KG}). Therefore to get higher order integrals of motion from the energy integral $E(v)=\int (\dot{v}^2+a^2(\nabla v)^2 +m^2 v^2)\,dx/2$ one can just  replace  $v$ by $\partial_x^\alpha v$. We get higher integrals of motion $I_\alpha (v)=E(\partial_x^\alpha v)$ for Eq. (\ref{KG}).  One expects that if $v\in H^{|\alpha|}$ then $J^\pm_\alpha (u)=I_\alpha(\Omega_\pm u)$
will be integrals of motion for Eq. (\ref{NLKG}). To construct higher order integrals of motion for the nonlinear equation one can use  also the Fourier transform as in the previous section.

Similar approach to construct higher order integrals of motion is applied also for the nonlinear Schr\"{o}dinger equation
$$
i\dot{u}+\bigtriangleup u+f(u)=0.
$$

It would be interesting to develop for nonlinear PDE an analogue of Koopman's approach to classical mechanics. To this end
we need a measure in an infinite-dimensional space which is invariant under the Hamilton flow and symplectic transformations.  A shift-invariant finitely additive measure
on a Hilbert space is constructed in \cite{Sak}.

\section{Wave operators and integrability in \\ quantum field theory}

Quantum field theory deals with operator-valued solutions of partial differential equations \cite{BOLOTO, GJ, Hepp}. For example, the scalar quantum field $\hat{u}=\hat{u}(t,x),\,t\in \mathbb{R}, x\in\mathbb{R}^n$ is an operator-valued function which should obey Eq. (\ref{NLKG})
$$
\hat{u}_{tt}-a^2\bigtriangleup \hat{u}+m^2\hat{ u}+f(\hat{u})=0,
$$
with initial data  satisfying the canonical commutation relation:
$$
[\hat{u}(0,x),\,\hat{u}_t(0,y)]=i\delta (x-y)
$$
A formal solution of these operator equations is given by the expression $\hat{u}(t,x)=e^{it\hat{H}}\hat{u}(0,x)e^{-it\hat{H}}$ where $\hat{H}$ is an operator obtained by quantization of Eq. (\ref{HNLKG})
\be\label{qh}
\hat{H}=\int_{\mathbb{R}^n}(\frac{1}{2}\hat{p}(x)^2 + \frac{1}{2}(\nabla \hat {q})^2 +          \frac{1}{2} m^2\hat{q}^2 +V(\hat {q}(x))dx 
\ee
and it is assumed that $\hat{u}(0,x)=\hat{q}(x),\,\hat{u}_t(0,x)=\hat{p}(x),\,[\hat{q}(x),\hat{p}(y)]=
i\delta(x-y)$.

After an appropriate regularization, one can construct the quantum field
$\hat{u}$ as an operator-valued distribution in a Fock Hilbert space ${\cal F}=\oplus_k L^2(\mathbb{R}^n)^{\otimes_s k}$.

The formal scattering theory in quantum field theory  is constructed analogously to the previous section.
There exists a  pair $(\hat{v}_+,\hat{v}_{-})$ of solutions (in and out fields) for the free Klein-Gordon equation
\begin{equation}\label{KGq}
\hat{v}_{tt}-a^2\bigtriangleup \hat{v}+m^2 \hat{v}=0,
\end{equation}
with initial data satisfying the relation
$$
[\hat{v}(0,x),\,\hat{v}_t(0,y)]=i\delta (x-y)
$$
such that the weak limit
$$
\lim_{t\to\pm\infty}(\hat{u}(t)-\hat{v}_{\pm}(t)=0.
$$
Moreover, the correspondence $(\hat{u}(0),\,\hat{u}_t(0))\mapsto (\hat{v}_{\pm}(0),\hat{v}_{\pm t}(0))$ defines the wave operators $\Omega_\pm$ on ${\cal F}$.

\section{Conclusions}
In this paper, it is noted that the Schr\"odinger equation with any self-adjoint Hamiltonian is unitarily equivalent to the set of classical harmonic oscillators and in this sense is completely integrable. This fact follows immediately from the spectral theorem which provides  the existence of the unitary transformation $W$ but not its explicit construction. The problem of how to construct the mapping $W$ is considered by using scattering theory and the wave operators. Higher order integrals of motion are indicated for a number of quantum and classical dynamical systems. Further study and clarification of the notion of complete integrability is desirable.

\section{Acknowledgements}
This work was initiated by  talks with V.V. Kozlov, who shared with the author  his ideas about quadratic integrals for the Schr\"odinger equation. The author is very grateful to I.Ya. Aref'eva, A.K. Gushchin, O. V. Inozemcev,  V.V. Kozlov, B.S. Mityagin, V.Zh. Sakbaev,  A.S. Trushechkin, B.O. Volkov
 and V.A. Zagrebnov for fruitful discussions.


\begin{thebibliography}{99}
\bibitem{AKN} V. I. Arnold, V. V. Kozlov, A. I. Neishtadt, {\it Mathematical aspects of classical and celestial mechanics}, Springer, 1997.


\bibitem{GGKM}
Clifford S. Gardner, John M. Green, Martin D. Kruskal, and Robert M. Miura, {\it Method for Solving the Korteweg-de Vries Equation},
Phys. Rev. Lett. 19, 1095 (1967).

\bibitem{FZ} V. E. Zakharov  and  L.D.Faddeev, {\it Korteweg–de Vries equation: a completely integrable Hamiltonian system}, Funct. Anal. Appl., 5:4 (1971), 280-287.

\bibitem{ZTF}  V. E. Zakharov,  A. Takhtadzhyan and L.D.Faddeev, "Complete description of solutions of the 'sine-Gordon' equation", Soviet Phys. Dokl., 19:12 (1974), 824-826

\bibitem{AK} Ya. Arefeva, V. Korepin,  {\it Scattering in two-dimensional model with Lagrangian $(1/\gamma) ((\partial_\mu u)^2/2 + m^2 (\cos u-1))$}, Pisma JETP Letters, 20:10 (1974), 680-684

\bibitem{Koz83} V. V. Kozlov, {\it Integrability and non-integrability in Hamiltonian mechanics}, Russian Math. Surveys, 38:1 (1983), 1-76.

\bibitem{NZ}  S. P. Novikov, V. E. Zakharov, S. V. Manakov  and L. P. Pitaevskii,          {\it Theory of Solitons:
The Inverse Scattering Method}, Springer, 1984.

\bibitem{TF} L. A. Takhtadzhyan, L. D. Faddeev, Hamiltonian methods in the theory of solitons, Nauka, Moscow, 1986.

 \bibitem{VV} V. S. Vladimirov, I. V. Volovich, {\it Local and nonlocal currents for nonlinear equations}, Theor. Math. Phys. 62 (1985) 1.


\bibitem{KT}  V. V. Kozlov, D. V. Treschev, {\it Polynomial Conservation Laws in Quantum Systems}, Theoret. and Math. Phys., 140:3 (2004), 1283-1298.



\bibitem{MPW} Willard Miller Jr., Sarah Post, Pavel Winternitz, {\it Classical and Quantum Superintegrability with Applications}, 	arXiv:1309.2694 [math-ph].

\bibitem{Koz18}  Valery V. Kozlov, {\it Linear Hamiltonian Systems: Quadratic Integrals, Singular Subspaces and Stability}, Regul. Chaotic Dyn., 23:1 (2018), 26-46.


\bibitem{TS} D. V. Treschev,  A. A. Shkalikov, {\it
On the Hamiltonian Property of Linear Dynamical Systems in Hilbert Space},
Mathematical Notes, 2017, 101:6, 1033?1039.


\bibitem {Koz} V.V. Kozlov, {\it Linear systems with a quadratic integral and the complete integrability of the Schr\"{o}dinger equation},  Uspekhi. Mat. Nauk, 75:5 (2019),189-190,

\bibitem{ALV} L. Accardi, Y. G. Lu, and I. Volovich, {\it Quantum Theory and Its Stochastic Limit} (Springer, Berlin, 2002).

\bibitem{APV} L Accardi, A. N. Pechen, I. V. Volovich, {\it Quantum stochastic
equation for the low density limit}, J. Phys. A: Math.
Gen. 35, 4889 (2002).

\bibitem{FSP} S. N. Filippov, G. N. Semin, A. N. Pechen, {\it Irreversible quantum dynamics for a system with gas environment in the low-density limit and the semiclassical collision model}, https://arxiv.org/abs/1908.11202.

\bibitem {Bur} J.-F. Burnol, {\it Scattering on the p-adic field and a trace formula,} International Mathematics Research Notices, Volume 2000, Issue 2, 2000, Pages 57-70.

\bibitem {VVZ} V. S. Vladimirov, I. V. Volovich and E.I. Zelenov, {\it p-Adic Analysis and Mathematical Physics}, World Scientific, Singapore, 1994.

\bibitem {30} 	B. Dragovich, A. Yu. Khrennikov, S. V. Kozyrev, I. V. Volovich, E. I. Zelenov, {\it p-Adic Mathematical Physics: The First 30 Years}, P-Adic Numbers, Ultrametric Anal. Appl., 9:2 (2017),  87-121.

\bibitem{Vol} Igor V. Volovich, {\it Complete Integrability of Quantum and Classical Dynamical Systems},  p-Adic Numbers, Ultrametric Analysis and Applications, 2019, Vol. 11, No. 4, pp. 328–334.

\bibitem{CM} P.R.~ Chernoff and  J.E. Marsden, {\it Properties of infinite dimensional Hamiltonian systems}, Lecture Notes in Mathematics, v.425.

\bibitem{KS} V.V.~ Kozlov and  O.G.~ Smolyanov,  {\it Hamiltonian aspects of quantum theory}, Dokl. Math., 85:3 (2012), 416-420.

\bibitem{Kuk} S.B.~ Kuksin, {\it  Hamiltonian PDEs}, Handbook of dynamical systems, 2006.
\bibitem{QC} Hans-Jorgen Stockmann, {\it Quantum Chaos: An Introduction}, (1999) Cambridge University Press.


\bibitem {DS} N. Dunford and
J. T. Schwartz, {\it Linear operators. Part II. Spectral theory.} Interscience, New York, 1963.

\bibitem{RS} M. Reed, B. Simon, {\it Methods of Modern Mathematical Physics, I, Functional analysis,} Elsevier, 1972.

\bibitem {Sim} B. Simon, {\it Wave operators for classical particle scattering,}
Commun.Math. Phys. (1971) 23: 37.

\bibitem {DG} J. Derezinski, C. Gerard, {\it Scattering Theory of Classical and Quantum N-Particle Systems,} Springer, 1997.

\bibitem{RS3} M. Reed, B. Simon, {\it Methods of Modern Mathematical Physics, III, Scattering theory,} Elsevier, 1979.

\bibitem{Yaf} D. Yafaev, {\it Lectures on scattering theory}, arXiv:math/0403213.
\bibitem{GV}  J. Ginibre and G. Velo, {\it Time decay of finite energy solutions of the nonlinear Klein-Gordon
and Schr\"{o}dinger equations}, Ann. Inst. Henri. Poincar'e, 43 (1985) 399-442.

\bibitem{Nak} K. Nakanishi, {\it Remarks on the energy scattering for nonlinear Klein-Gordon and Schr\"{o}dinger
equations}, Tohoku Math. J. (2) 53 (2001), no. 2, 285-303.

\bibitem{ZJ}  Zihua Guo, Jia Shen, {\it A note on large data scattering for 2D non-linear Schrodinger and Klein-Gordon equation},  arXiv:1906.01804.

\bibitem{Sak} V.Zh. Sakbaev, {\it Averaging of random walks and shift-invariant measures on a Hilbert space},  Theoretical and Mathematical Physics, 2017, 191:3, 886-909.
\bibitem{BOLOTO}  N.N.~Bogolubov, A.A.~Logunov, A.I.~Oksak, I.~Todorov, {\it General Principles of Quantum Field Theory},
Springer Verlag, 2012.
\bibitem{GJ} J.~Glimm, A.~Jaffe, {\it Quantum Physics: A Functional Integral Point of View,}
Springer Verlag, 1987.
\bibitem{Hepp} K. Hepp, {\it Th\'eorie de la renormalisation},  Lecture Notes in  Physics, 1969.

\end{thebibliography}
\end{document}